# Non-linear index coding outperforming the linear optimum

Eyal Lubetzky[*]  Uri Stav[†]

**Abstract**

The following source coding problem was introduced by Birk and Kol: a sender holds a word $x \in \{0,1\}^n$, and wishes to broadcast a codeword to $n$ receivers, $R_1, \ldots, R_n$. The receiver $R_i$ is interested in $x_i$, and has prior *side information* comprising some subset of the $n$ bits. This corresponds to a directed graph $G$ on $n$ vertices, where $ij$ is an edge iff $R_i$ knows the bit $x_j$. An *index code* for $G$ is an encoding scheme which enables each $R_i$ to always reconstruct $x_i$, given his side information. The minimal word length of an index code was studied by Bar-Yossef, Birk, Jayram and Kol [4]. They introduced a graph parameter, $\mathrm{minrk}_2(G)$, which completely characterizes the length of an optimal *linear* index code for $G$. The authors of [4] showed that in various cases linear codes attain the optimal word length, and conjectured that linear index coding is in fact *always* optimal.

In this work, we disprove the main conjecture of [4] in the following strong sense: for any $\varepsilon > 0$ and sufficiently large $n$, there is an $n$-vertex graph $G$ so that every linear index code for $G$ requires codewords of length at least $n^{1-\varepsilon}$, and yet a non-linear index code for $G$ has a word length of $n^{\varepsilon}$. This is achieved by an explicit construction, which extends Alon's variant of the celebrated Ramsey construction of Frankl and Wilson.

In addition, we study optimal index codes in various, less restricted, natural models, and prove several related properties of the graph parameter $\mathrm{minrk}(G)$.

## 1 Introduction

Source coding deals with a scenario in which a *sender* has some data string $x$ he wishes to transmit through a broadcast channel to *receivers*. The first and classical result in this area is Shannon's Source Coding Theorem. This has been followed by various scenarios which differ in the nature of the data to be transmitted, the broadcast channel and some assumptions on the computational abilities of the users. Another family of source coding problems, which attracted a considerable amount of attention over the years, deals with the assumption that the receivers possess some prior knowledge on the data string $x$. It was shown that in some cases even some restricted assumptions on this knowledge may drastically affect the nature of the coding problem.

In this paper we consider a variant of source coding which was first proposed by Birk and Kol [5]. In this variant, called Informed Source Coding On Demand (ISCOD), each receiver has some prior side information, comprising some subset of the input word $x$. The sender is aware of the portion of $x$ known to each receiver. Moreover, each receiver is interested in just part of the data. Following [4], we restrict ourselves to the problem which is formalized as follows.

---

[*]Microsoft Research, One Microsoft Way, Redmond, WA 98052-6399, USA. Email: eyal@microsoft.com.

[†]School of Computer Science, Raymond and Beverly Sackler Faculty of Exact Sciences, Tel Aviv University, Tel Aviv, 69978, Israel. Email: uristav@tau.ac.il.

**Definition 1** (**index code**). *A sender wishes to send a word $x \in \{0,1\}^n$ to $n$ receivers $R_1, \ldots, R_n$. Each $R_i$ knows some of the bits of $x$ and is interested solely in the bit $x_i$. An* index code *of length $\ell$ for this setting is a binary code of word-length $\ell$, which enables $R_i$ to recover $x_i$ for any $x$ and $i$.*

Using a graph model for the side-information, this problem can be restated as a graph parameter. For a directed graph $G$ and a vertex $v$, let $N_G^+(v)$ be the set of out-neighbors of $v$ in $G$, and for $x \in \{0,1\}^n$ and $S \subset [n] = \{1, \ldots, n\}$, let $x|_S$ be the restriction of $x$ to the coordinates of $S$.

**Definition 2** ($\ell(\mathbf{G})$). *The setting of Definition 1 is characterized by the directed* side information graph $G$ *on the vertex set $[n]$, where $(i, j)$ is an edge iff $R_i$ knows the value of $x_j$. An* index code *of length $\ell$ for $G$ is a function $E : \{0, 1\}^n \to \{0, 1\}^\ell$ and functions $D_1, \ldots, D_n$, so that for all $i \in [n]$ and $x \in \{0, 1\}^n$, $D_i(E(x), x|_{N_G^+(i)}) = x_i$. Denote the minimal length of an index code for $G$ by $\ell(G)$.*

**Example:** Suppose that every receiver $R_i$ knows in advance the whole word $x$, except for the single bit $x_i$ he wishes to recover. The corresponding side information graph $G$ is the complete graph $K_n$ (that is, $(i,j)$ is an edge for all $i \neq j$). By broadcasting the XOR of all the bits of $x$, each receiver can easily compute its missing bit:

$$E(x) = \bigoplus_{i=1}^n x_i \ ,$$
$$D_i(E(x), x|_{\{j:j\neq i\}}) = E(x) \oplus (\bigoplus_{j \neq i} x_j) = x_i \ .$$

In this case the code has length $\ell = 1$ and $E$ is a linear function of $x$ over $GF(2)$.

The problem of Informed Source Coding On Demand (ISCOD) was presented by Birk and Kol [5]. They were motivated by various applications of distributed communication such as satellite communication networks with caching clients. In such applications, the clients have limited storage and maintain part of the transmitted information. Subsequently, the clients receive requests for arbitrary information blocks, and may use a slow backward channel to advise the server of their status. The server, playing the role of the sender in Definition 1, then broadcasts a single transmission to all clients (the receivers). As observed by Birk and Kol [5], when the sender has only partial knowledge of the side information (e.g., the *number* of missing blocks for each user), an erasure correcting code such as Reed-Solomon Code performs well. This is also the case if every user is expected to be able to decode the whole information. The authors of [5] present some bounds and heuristics for obtaining efficient encoding schemes, as well as protocols for implementing the above scenario. See [5] and [4] for more details on the relation between the source coding problem, as formulated above, and the ISCOD problem, as well as the communication complexity of the indexing function, random access codes and network coding.

Bar-Yossef, Birk, Jayram and Kol [4] further investigated index coding. They showed that this problem is different in nature from the well-known source coding problems previously studied by Witsenhausen in [12]. Their main contribution is an upper bound on $\ell(G)$, the optimal length of an index code (Definition 2). The upper bound is a graph parameter denoted by $\mathrm{minrk}_2(G)$, which is also shown to be the length of the optimal *linear* index code. It is shown in [4] that in several cases linear codes are in fact optimal, e.g., for directed acyclic graphs, perfect graphs, odd cycles and odd anti-holes. An information theoretic lower bound on $\ell(G)$ is obtained: it is at least the size of a maximal acyclic induced subgraph of $G$. This lower bound holds even for the relaxed problem of *randomized* index codes, where the sender is allowed to use (public) random coins during encoding, and the receivers are expected to decode their information correctly with high probability over these coin flips. Nevertheless, they show that in some cases the lower bound is not tight.

Having proved that the upper bound $\ell(G) \leq \mathrm{minrk}_2(G)$ is tight for several natural graph families and under some relaxed restrictions on the code ("semi-linearly-decodable"), the authors of [4] conjectured that the length of the optimal index code is in fact equal to $\mathrm{minrk}_2(G)$. That is, they conjectured that linear index coding is always optimal, and concluded that this was the main open problem to be investigated.

Before stating the main results of this paper, we review the definition of $\mathrm{minrk}_2(G)$ and other related graph theoretic parameters.

## 1.1 Definitions, notations and background

Let $G = (V, E)$ be a directed graph on the vertex set $V = [n]$. The adjacency matrix of $G$, denoted by $A_G = (a_{ij})$, is the $n \times n$ binary matrix where $a_{ij} = 1$ iff $(i, j) \in E$. An *independent set* of $G$ is a set of vertices which have no edges between them, and the *independence number* of $G$, $\alpha(G)$, is the cardinality of a maximum independent set. The *chromatic number* of $G$, $\chi(G)$, is the minimum number of independent sets whose union is all of $V$. Let $\overline{G}$ denote the *graph complement* of $G$. A *clique* of $G$ is an independent set of $\overline{G}$ (i.e., a set of vertices such that all edges between them belong to $G$), and the *clique number* of $G$, $\omega(G)$, is the cardinality of a maximum clique. Without being formal, a graph $G$ is called "Ramsey" if both $\alpha(G)$ and $\omega(G)$ are "small".

In [4], a binary $n \times n$ matrix $A = (a_{ij})$ was said to "fit" $G$ if $A$ has 1-s on its diagonal, and 0 in all the indices $i, j$ where $i \neq j$ and $(i, j) \notin E$. The parameter $\mathrm{minrk}_2(G)$ was defined to be the minimal possible rank over $GF(2)$ of a matrix which fits $G$.

To extend this definition to a general field, let $A = (a_{ij})$ be an $n \times n$ matrix over some field $\mathbb{F}$. We say that $A$ *represents* the graph $G$ over $\mathbb{F}$ if $a_{ii} \neq 0$ for all $i$, and $a_{ij} = 0$ whenever $i \neq j$ and $(i, j) \notin E$. The *minrank* of a directed graph $G$ with respect to the field $\mathbb{F}$ is defined by

$$\mathrm{minrk}_{\mathbb{F}}(G) = \min\{\mathrm{rank}_{\mathbb{F}}(A) : A \text{ represents } G \text{ over } \mathbb{F}\} .$$

For the common case where $\mathbb{F}$ is a finite field, we abbreviate:

$$\mathrm{minrk}_{p^k}(G) = \mathrm{minrk}_{GF(p^k)}(G) .$$

The notion of $\mathrm{minrk}(G)$ for an undirected graph $G$ was first considered in the context of graph capacities by Haemers [8],[9]. The Shannon capacity of the graph $G$, denoted by $c(G)$, is a notoriously challenging parameter, which was defined by Shannon in [11], and remains unknown even for simple graphs, such as $C_7$, the cycle on 7 vertices. Lower bounds for $c(G)$ are given in terms of independence numbers of certain graphs, and in particular, $\alpha(G) \leq c(G)$. Haemers showed that for all $\mathbb{F}$, $\mathrm{minrk}_{\mathbb{F}}(G)$ is sandwiched between $c(G)$ and $\chi(\overline{G})$, the chromatic number of the complement of $G$, altogether giving

$$\alpha(G) \leq c(G) \leq \mathrm{minrk}_{\mathbb{F}}(G) \leq \chi(\overline{G}) . \tag{1}$$

While $\mathrm{minrk}_\mathbb{F}(G)$ can prove to be difficult to compute, the most useful upper bound for $c(G)$ is $\vartheta(G)$, the Lovász $\vartheta$-function, which was introduced in the seminal paper [10] to compute $c(C_5)$. The matrix-rank argument was thereafter introduced by Haemers to answer some questions of [10], and has since been used (under some variants) in additional settings to obtain better bounds than those provided by the $\vartheta$-function (cf., e.g., [1]).

## 1.2 New results

The main result of this paper is an improved index coding scheme, which is shown to strictly improve upon the $\mathrm{minrk}_2(G)$ bound. This disproves the main conjecture of [4] regarding the optimality of linear index coding, as stated by the following theorem.

**Theorem 1.1.** *For any $\varepsilon > 0$ and any sufficiently large $n$, there is an $n$-vertex graph $G$ so that:*

1. *Any linear index code for $G$ requires $n^{1-\varepsilon}$ bits, that is, $\mathrm{minrk}_2(G) \geq n^{1-\varepsilon}$.*

2. *There exists a non-linear index code for $G$ using $n^\varepsilon$ bits, that is, $\ell(G) \leq n^\varepsilon$.*

*Moreover, the graph $G$ is undirected and can be constructed explicitly.*

Note that this in fact disproves the conjecture of Bar-Yossef et al. in the following strong sense: the ratio between an optimal code and an optimal linear code over $GF(2)$ can be $n^{1-o(1)}$. The essence of the proof lies in the fact that, in some cases, linear codes over higher-order fields may yield significantly better index coding schemes. The term "linear codes over $GF(p)$" is used here to describe a coding scheme, in which the input word is encoded into a sequence of linear functionals of its symbols over $GF(p)$, which are subsequently used for the decoding (the protocol for transmitting these functionals need not be linear). However, as the next theorem demonstrates, even this extended family of index codes may be suboptimal.

**Theorem 1.2.** *For any $\varepsilon > 0$ and any sufficiently large $n$, there is an $n$-vertex graph $G$ so that:*

1. *Any linear index code for $G$ over some field $\mathbb{F}$ requires $\sqrt{n}$ symbols, that is, $\mathrm{minrk}_\mathbb{F}(G) \geq \sqrt{n}$.*

2. *There exists a non-linear index code for $G$ using $n^\varepsilon$ bits, that is, $\ell(G) \leq n^\varepsilon$.*

*Moreover, the graph $G$ is undirected and can be constructed explicitly.*

In order to prove the above two theorems, we introduce the following upper bound on $\ell(G)$, which is a simple extension of a result of [4] (the special case $\mathbb{F} = GF(2)$), and is a special case of Proposition 2.1 (Section 2).

$$\ell(G) \leq \min_{\mathbb{F} \,:\, |\mathbb{F}| < \infty} \left\lceil \mathrm{minrk}_\mathbb{F}(G) \log_2 |\mathbb{F}| \right\rceil . \qquad (2)$$

The proof of Theorem 1.1 relies on the fact that for some graphs, the minimum of (2) is attained when $\mathbb{F} \neq GF(2)$, in which case the linear code over $GF(2)$ is suboptimal. Proposition 2.2 (Section 2) provides a construction of such graphs, and is the main ingredient in the proof of Theorem 1.1. This proposition, which may be of independent interest, states that for any pair of finite fields with distinct characteristics, $\mathbb{F}$ and $\mathbb{K}$, the gap between $\text{minrk}_{\mathbb{F}}$ and $\text{minrk}_{\mathbb{K}}$ can be $n^{1-o(1)}$. Theorem 1.1 is then obtained as a corollary of (2) and a special case of Proposition 2.2.

Moreover, as Theorem 1.2 shows, the upper bound of (2) is not always tight. To see this, we combine the construction in the above mentioned Proposition 2.2 with some additional ideas.

As an additional corollary, Proposition 2.2 yields that $\text{minrk}_{\mathbb{F}}(G)/\vartheta(G)$ (where $\vartheta$ is the Lovász $\vartheta$-function and $|V(G)| = n$) is in some cases (roughly) at least $\sqrt{n}$, whereas in other cases it is (roughly) at most $1/\sqrt{n}$. This addresses another question of [4] on the relation between these two parameters. The relation between $\ell(G)$ and the Shannon capacity of $G$, $c(G)$, is addressed as well, as a by-product of the proof of Theorem 1.2.

We also extend the main construction of Proposition 2.2 and give, for any prescribed set of finite fields $\{\mathbb{F}_i\}$ and an additional finite field $\mathbb{K}$ of a distinct characteristic, a construction of a graph $G$ so that $\text{minrk}_{\mathbb{F}_i}(G)$ is "large" for all $i$, whereas $\text{minrk}_{\mathbb{K}}(G)$ is "small".

**Proposition 1.3.** *For any fixed $t$, let $\mathbb{F}_1, \ldots, \mathbb{F}_t$ denote finite fields, and let $\mathbb{K}$ denote a finite field of a distinct characteristic. For any $\varepsilon > 0$ and a sufficiently large $n$, there is an explicit construction of a graph $G$ on $n$ vertices, so that $\text{minrk}_{\mathbb{K}}(G) \leq n^{\varepsilon}$, whereas for all $i \in [t]$, $\text{minrk}_{\mathbb{F}_i}(G) \geq n^{1-\varepsilon}$.*

In the second part of this paper, we revisit the problem definition. It is shown that the restricted problem given in Definition 1 captures many other cases arising from the original distributed applications, which motivated the study of Informed Source Coding On Demand. In particular, we suggest appropriate models and reductions for cases in which multiple users are interested in the same bit, there are multiple rounds of transmission and the transmitted words are over a large alphabet. These models are obtained as natural extensions of the original problem, and exhibit interesting relations to the parameters $\ell(G)$ and $\text{minrk}(G)$.

## 1.3 Techniques

A key element in the proof of the main result is an extended version of the Ramsey graph constructed by Alon [1], which is a variant of the well-known Ramsey construction of Frankl and Wilson [7]. This graph, $G_{p,q}$ for some large primes $p, q$, was used by Alon in order to disprove an old conjecture of Shannon [11] on the Shannon capacity of a union of graphs.

Using some properties of the minrk parameter, one can show that the graph $G_{p,q}$ has a "small" $\text{minrk}_p$ and a "large" $\text{minrk}_q$, implying that the optimal linear index code over $GF(p)$ may be significantly better than the one over $GF(q)$. However, it is imperative in the above construction that both $p$ and $q$ will be large, whereas we are interested in the case $q = 2$, corresponding to $\text{minrk}_2$. To this end, we extend the above construction of [1] to prime-powers, using some classical results on congruencies of binomial coefficients. This allows omitting the requirement that $p, q$ should be large, by taking sufficiently large powers of *arbitrary* distinct primes $p$ and $q$.

Using variants of the above construction, we extend the results to multiple fields, to obtain Theorem 1.2 and Proposition 1.3. En route, we derive several properties of the minrank parameter, which may be of independent interest.

The proofs of the results throughout the paper combine arguments from Linear Algebra and Number Theory along with some additional ideas, inspired by the theory of graph capacities under the strong graph product definition.

## 1.4 Organization

The rest of the paper is organized as follows. Section 2 contains a description of the basic construction, and the proof of Theorem 1.1. The extension of this result to multiple fields, including the proof of Theorem 1.2, appears in Section 3. In Section 4, we study the various extensions of the original problem. Section 5 contains some concluding remarks and open problems.

## 2 Linear index codes over higher-order fields

In this section, we prove Theorem 1.1, by constructing graphs for which a given linear index code over a higher-order field outperforms all linear index codes over $GF(2)$.

### 2.1 Proof of Theorem 1.1

The first ingredient in the proof is a linear index coding scheme, which is an extension of the ideas in [4] for larger fields. This notion is formulated in the next proposition, whose proof appears in Subsection 2.2.

**Proposition 2.1.** *Let $G$ be a graph, and let $A$ be a matrix which represents $G$ over some field $\mathbb{F}$ (not necessarily finite). Then $\ell(G) \leq \lceil \log_2 |\{Ax : x \in \{0,1\}^n\}| \rceil$. In particular, the following holds:*
$$\ell(G) \leq \min_{\mathbb{F}\,:\,|\mathbb{F}|<\infty} \lceil \operatorname{minrk}_{\mathbb{F}}(G) \log_2 |\mathbb{F}| \rceil .$$

The second and main ingredient in the proof of Theorem 1.1 is Proposition 2.2, whose proof appears in Subsection 2.3. Here and in what follows, all logarithms are in the natural base unless stated otherwise.

**Proposition 2.2.** *Let $\mathbb{F}$ and $\mathbb{K}$ denote two finite fields with distinct characteristics. There is an explicit construction of a family of graphs $G = G(n)$ on $n$ vertices, so that*
$$\operatorname{minrk}_{\mathbb{F}}(G) \leq \exp\left(\sqrt{(2+o(1))\log n \log\log n}\right)$$
$$= n^{o(1)} , \qquad (3)$$
*and yet:*
$$\operatorname{minrk}_{\mathbb{K}}(G) \geq n/\exp\left(\sqrt{(2+o(1))\log n \log\log n}\right)$$
$$= n^{1-o(1)} , \qquad (4)$$
*where the $o(1)$-terms tend to $0$ as $n \to \infty$.*

In order to derive Theorem 1.1 from Propositions 2.1 and 2.2, apply Proposition 2.2, setting $\mathbb{F} = GF(p)$ and $\mathbb{K} = GF(2)$, where $p > 2$ is any fixed (odd) prime. Let $\varepsilon > 0$; for any sufficiently large $n$, the graph obtained above satisfies
$$\operatorname{minrk}_2(G) \geq n/\exp(O(\sqrt{\log n \log\log n})) \geq n^{1-\varepsilon} ,$$

and hence, by the results of [4], any linear index code over $GF(2)$ requires a word length of at least $n^{1-\varepsilon}$ bits. On the other hand, Proposition 2.1 implies that

$$\begin{aligned} \ell(G) &\leq \lceil \mathrm{minrk}_p(G) \log_2(p) \rceil \\ &\leq \exp(O(\sqrt{\log n \log \log n})) \leq n^\varepsilon \;. \end{aligned}$$

∎

## 2.2 Proof of Proposition 2.1

Let $V = [n]$ denote the vertex set of $G$, $A = (a_{ij})$ denote a matrix which represents $G$ over some field $\mathbb{F}$ (not necessarily finite), and $S = \{Ax : x \in \{0,1\}^n\} \subset \mathbb{F}^n$. For some arbitrary ordering of the elements of $S$, the encoding of $x \in \{0,1\}^n$ is the label of $Ax$, requiring a word-length of $\lceil \log_2 |S| \rceil$ bits. For decoding, the $i$-th receiver $R_i$ examines $(Ax)_i$, and since the diagonal of $A$ does not contain zero entries by definition, we have:

$$\begin{aligned} a_{ii}^{-1}(Ax)_i &= a_{ii}^{-1} \sum_j a_{ij} x_j \\ &= x_i + a_{ii}^{-1} \sum_{j \in N_G^+(i)} a_{ij} x_j \;, \end{aligned} \quad (5)$$

where the last equality is by the fact that $A$ represents $G$. As $R_i$ knows $\{x_j : j \in N_G^+(i)\}$, this allows $R_i$ to recover $x_i$. Therefore, indeed $\ell(G) \leq \lceil \log_2 |S| \rceil$.

To conclude the proof, note that in case $\mathbb{F}$ is finite, we have $|S| \leq |\mathbb{F}|^{\mathrm{rank}_\mathbb{F}(A)}$, as required. Furthermore, in this case it is possible to use a linear code utilizing the same word-length. The sender transmits a binary-encoding of the inner-products $(u_1 \cdot x, \ldots, u_r \cdot x) \in \mathbb{F}^r$, where $\{u_1, \ldots, u_r\}$ is a basis for the rows of $A$ over $\mathbb{F}$. ∎

**Remark 2.3:** As proved for the case $\mathbb{F} = GF(2)$ in [4], it is possible to show that the above bound is tight for the case of linear codes over $\mathbb{F}$. That is, the length of an optimal linear index code over a finite field $\mathbb{F}$ is $\lceil \mathrm{minrk}_\mathbb{F} \log_2 |\mathbb{F}| \rceil$.

## 2.3 Proof of Proposition 2.2

We first consider the case $\mathbb{F} = GF(p)$ and $\mathbb{K} = GF(q)$ for distinct primes $p$ and $q$. Let $\varepsilon > 0$, and let $k$ denote a (large) integer satisfying[1]

$$q^l < p^k < (1+\varepsilon) q^l \;, \quad \text{where } l = \lfloor k \log_q p \rfloor \;. \quad (6)$$

Define:

$$s = p^k q^l - 1 \quad \text{and} \quad r = p^{3k} \;. \quad (7)$$

---

[1] It is easy to verify that there are infinitely many such integers $k$, as $p, q$ are distinct primes, and hence the set $\{k \log_q p \pmod 1\}_{k \in \mathbb{N}}$ is dense in $[0, 1]$.

The graph $G$ on $n = \binom{r}{s}$ vertices[2] is defined as follows. Its vertices are all $s$-element subsets of $[r]$, and two vertices are adjacent iff their corresponding sets have an intersection whose cardinality is congruent to $-1$ modulo $p^k$:

$$V(G) = \binom{[r]}{s}, \tag{8}$$

$$(X,Y) \in E(G) \iff \begin{cases} X \neq Y, \\ |X \cap Y| \equiv -1 \pmod{p^k}. \end{cases}$$

For some integer $d$ to be determined later, define the *inclusion matrix* $M_d$ to be the $\binom{r}{s} \times \binom{r}{d}$ binary matrix, indexed by all $s$-element and $d$-element subsets of $[r]$, where $(M_d)_{A,B} = 1$ iff $B \subset A$, for all $A \in \binom{[r]}{s}$ and $B \in \binom{[r]}{d}$. Notice that the $n \times n$ matrix $M_d(M_d)^T$ satisfies the following for all $A, B \in V$ (not necessarily distinct):

$$(M_d(M_d)^T)_{A,B} = \left| \left\{ X \in \binom{[r]}{d} : X \subset (A \cap B) \right\} \right|$$
$$= \binom{|A \cap B|}{d}. \tag{9}$$

Define $P = M_{p^k-1}(M_{p^k-1})^T$ and $Q = M_{q^l-1}(M_{q^l-1})^T$. We claim that $P$ represents $G$ over $GF(p)$ whereas $Q$ represents $\overline{G}$ over $GF(q)$. To see this, we need the following simple observation, which is a special case of Lucas's Theorem (cf., e.g., [6]) on congruencies of binomial coefficients. It was used, for instance, in [3] for constructing low-degree representations of OR functions modulo composite numbers, as well as in [7].

**Observation 2.4.** *For every prime $p$ and integers $i, j, e$ with $i < p^e$, $\binom{j + p^e}{i} \equiv \binom{j}{i} \pmod{p}$.*

Consider some $A \in V$; since $s$ satisfies both $s \equiv (p^k - 1) \pmod{p^k}$ and $s \equiv (q^l - 1) \pmod{q^l}$, combining (9) with Observation 2.4 gives

$$P_{A,A} = \binom{s}{p^k - 1} \equiv 1 \pmod{p},$$

and

$$Q_{A,A} = \binom{s}{q^l - 1} \equiv 1 \pmod{q}.$$

Thus, indeed the diagonal entries of $P$ and $Q$ are non-zero; it remains to show that their $(A, B)$-entries are 0 wherever $A, B$ are distinct non-adjacent vertices. To this end, take $A, B \in V$ so that $A \neq B$ and $AB \notin E(G)$; by (8), $|A \cap B| \not\equiv -1 \pmod{p^k}$, hence

$$P_{A,B} = \binom{|A \cap B|}{p^k - 1} \equiv 0 \pmod{p},$$

the last equivalence again following from Observation 2.4, as $\binom{x}{p^k-1} = 0$ for all $x \in \{0, \ldots, p^k - 2\}$. Finally, suppose that $A, B \in V$ satisfy $A \neq B$ and $AB \notin E(\overline{G})$. That is, $AB \in E(G)$, hence $|A \cap B| \equiv -1 \pmod{p^k}$. The Chinese Remainder Theorem now implies $|A \cap B| \not\equiv -1 \pmod{q^l}$, otherwise we would get $|A \cap B| = s$ and $A = B$. Since $\binom{x}{q^l-1} = 0$ for all $x \in \{0, \ldots, q^l - 2\}$, we get

$$Q_{A,B} = \binom{|A \cap B|}{q^l - 1} \equiv 0 \pmod{q} .$$

---

[2] By well known properties of the density of prime numbers, and standard graph theoretic arguments, proving the assertion of the proposition for these values of $n$ in fact implies the result for any $n$.

Altogether, $P$ represents $G$ over $GF(p)$, and $Q$ represents $\overline{G}$ over $GF(q)$. Therefore, $\mathrm{minrk}_p(G)$ is at most $\mathrm{rank}_p(P) \leq \mathrm{rank}_p(M_{p^k-1})$, and similarly, $\mathrm{minrk}_q(\overline{G})$ is at most $\mathrm{rank}_q(Q) \leq \mathrm{rank}_q(M_{q^l-1})$. As $M_d$ has $\binom{r}{d}$ columns, $n = \binom{p^{3k}}{p^k q^l}$ and $q^l < p^k < (1+\varepsilon)q^l$, a straightforward calculation now gives:

$$\mathrm{minrk}_p(G) \leq \binom{r}{p^k - 1}$$
$$< \exp\left(\sqrt{(1 + \varepsilon + o(1))2 \log n \log \log n}\right) ,$$
$$\mathrm{minrk}_q(\overline{G}) \leq \binom{r}{q^l - 1}$$
$$< \exp\left(\sqrt{(1 + \varepsilon + o(1))2 \log n \log \log n}\right) .$$

The next simple claim relates $\mathrm{minrk}_q(G)$ and $\mathrm{minrk}_q(\overline{G})$:

**Claim 2.5.** *For any graph $G$ on $n$ vertices and any field $\mathbb{F}$, $\mathrm{minrk}_\mathbb{F}(G) \cdot \mathrm{minrk}_\mathbb{F}(\overline{G}) \geq n$.*

*Proof.* We use the following definition of graph product due to Shannon [11]: $G_1 \times G_2$, the *strong graph product* of $G_1$ and $G_2$, is the graph whose vertex set is $V(G_1) \times V(G_2)$, where two distinct vertices $(u_1, u_2) \neq (v_1, v_2)$ are adjacent iff for all $i \in \{1, 2\}$, either $u_i = v_i$ or $(u_i, v_i) \in E(G_i)$.

As observed by Haemers [8], if $A_1$ and $A_2$ represent $G_1$ and $G_2$ respectively over $\mathbb{F}$, then the tensor product $A_1 \otimes A_2$ represents $G_1 \times G_2$ over $\mathbb{F}$. To see this, notice that the diagonal of $A_1 \otimes A_2$ does not contain zero entries, and that if $(u_1, u_2) \neq (v_1, v_2)$ are disconnected vertices in $G_1 \times G_2$, then by definition $(A_1)_{(u_1,v_1)}(A_2)_{(u_2,v_2)} = 0$, since in this case for some $i \in \{1, 2\}$ we have $u_i \neq v_i$ and $u_i v_i \notin E(G_i)$. Letting $A_1$ and $A_2$ denote matrices which attain $\mathrm{minrk}_\mathbb{F}(G)$ and $\mathrm{minrk}_\mathbb{F}(\overline{G})$ respectively, the above discussion implies that:

$$\mathrm{minrk}_\mathbb{F}(G \times \overline{G}) \leq \mathrm{rank}(A_1 \otimes A_2)$$
$$= \mathrm{minrk}_\mathbb{F}(G) \cdot \mathrm{minrk}_\mathbb{F}(\overline{G}) .$$

However, the set $\{(u, u) : u \in V(G)\}$ is an independent-set of $G \times \overline{G}$, since for $u \neq v$, either $uv \in E(G)$ and $uv \notin E(\overline{G})$ or vice versa. Therefore, (1) gives $\mathrm{minrk}_\mathbb{F}(G \times \overline{G}) \geq \alpha(G \times \overline{G}) \geq n$, completing the proof of the claim. ∎

This concludes the proof of the proposition for the case $\mathbb{F} = GF(p)$, $\mathbb{K} = GF(q)$, where $p, q$ are two distinct primes. The generalization to the case of prime-powers is an immediate consequence of the next claim:

**Claim 2.6.** *Let $G$ be a graph, $p$ be a prime and $k$ be an integer. The following holds:*

$$\frac{1}{k} \mathrm{minrk}_p(G) \leq \mathrm{minrk}_{p^k}(G) \leq \mathrm{minrk}_p(G) . \tag{10}$$

*Proof.* The statement $\mathrm{minrk}_{p^k}(G) \leq \mathrm{minrk}_p(G)$ follows immediately from the fact that any matrix $A$ which represents $G$ over $GF(p)$ also represents $G$ over $GF(p^k)$, and in addition satisfies $\mathrm{rank}_{p^k}(A) \leq \mathrm{rank}_p(A)$.

To show that $\mathrm{minrk}_p(G) \leq k\,\mathrm{minrk}_{p^k}(G)$, let $V = [n]$ denote the vertex set of $G$, and let $A = (a_{ij})$ denote a matrix which represents $G$ over $GF(p^k)$ with rank $r = \mathrm{minrk}_{p^k}(G)$. As usual, we represent the elements of $GF(p^k)$ as polynomials of degree at most $k-1$ over $GF(p)$ in the variable $x$. Since the result of multiplying each row of $A$ by a non-zero element of $GF(p^k)$ is a matrix of rank $r$ which also represents $G$ over $GF(p^k)$, assume without loss of generality that $a_{ii} = 1$ for all $i \in [n]$. By this assumption, the $n \times n$ matrix $B = (b_{ij})$, which contains the free coefficients of the polynomials in $A$, represents $G$ over $GF(p)$. To complete the proof, we claim that $\mathrm{rank}_p(B) \leq kr$. This follows from the simple fact that, if $\{u_1, \ldots, u_r\}$ is a basis for the rows of $A$ over $GF(p^k)$, then the set $\bigcup_{i=1}^{r}\{u_i, x \cdot u_i, \ldots, x^{k-1} \cdot u_i\}$ spans the rows of $A$ when viewed as $kn$-dimensional vectors over $GF(p)$. ∎

This concludes the proof of Proposition 2.2. ∎

**Remark 2.7:** Alon's Ramsey construction [1] is the graph on the vertex set $V = \binom{[r]}{s}$, where $r = p^3$ and $s = pq - 1$ for some large primes $p \sim q$, and two distinct vertices $A, B$ are adjacent iff $|A \cap B| \equiv -1 \pmod{p}$. Our construction allows $p$ and $q$ to be large prime-powers $p^k \sim q^l$. Note that the original construction by Frankl and Wilson [7] had the parameters $r = q^3$ and $s = q^2 - 1$ for some prime-power $q$, and two distinct vertices $A$ and $B$ are adjacent iff $|A \cap B| \equiv -1 \pmod{q}$.

**Remark 2.8:** Another corollary of Proposition 2.2 is that the ratio between $\mathrm{minrk}_{\mathbb{F}}(G)$ and $\vartheta(G)$ can be arbitrarily large. To see this, consider the $n$-vertex graph $G$ constructed in Proposition 2.2 for $\mathbb{F} = GF(p)$ and $\mathbb{K} = GF(q)$, where $p$ and $q$ are two distinct primes: it satisfies $\mathrm{minrk}_p(G) \leq n^{o(1)}$ and $\mathrm{minrk}_q(\overline{G}) \leq n^{o(1)}$. Clearly, $G$ is vertex transitive (that is, its automorphism group is closed under all vertex substitutions), as we can always relabel the elements of the ground set $[r]$. By [10] (Theorem 9), every vertex transitive graph $G$ on $n$ vertices satisfies

$$\vartheta(G)\vartheta(\overline{G}) = n \ .$$

Assume without loss of generality that $\vartheta(G) \geq \sqrt{n} \geq \vartheta(\overline{G})$ (otherwise, switch the roles of $p$ and $q$ and of $G$ and $\overline{G}$). As $\mathrm{minrk}_p(G) \leq n^{o(1)}$ and $\mathrm{minrk}_p(\overline{G}) \geq n^{1-o(1)}$, we deduce that

$$\vartheta(G) \geq n^{\frac{1}{2}-o(1)} \cdot \mathrm{minrk}_p(G) \ ,$$

and yet

$$\mathrm{minrk}_p(\overline{G}) \geq n^{\frac{1}{2}-o(1)} \cdot \vartheta(\overline{G}) \ .$$

## 3 Outperforming linear index codes over multiple fields

In this section we use variants of the graphs constructed in Proposition 2.2 in order to prove Theorem 1.2 and Proposition 1.3.

## 3.1 Proof of Theorem 1.2

Let $\varepsilon > 0$, and let $G$ be the graph constructed by Proposition 2.2 for $\mathbb{F} = GF(2)$, $\mathbb{K} = GF(3)$, and a sufficiently large $n$ such that $\mathrm{minrk}_2(G) \leq n^{\varepsilon/2}$ and $\mathrm{minrk}_3(\overline{G}) \leq n^{\varepsilon/2}$. Let $H$ denote the graph $G + \overline{G}$, that is, the disjoint union of $G$ and its complement. We claim that

$$\ell(H) < 3n^{\varepsilon/2} \text{ , and yet } c(H) \geq \sqrt{2n} = \sqrt{|V(H)|} \text{ .}$$

To see this, observe that in order to obtain an index code for a given graph, one may always arbitrarily partition the graph into subgraphs and concatenate their individual index codes:

**Observation 3.1.** *For any graph $G$ and any partition of $G$ to subgraphs $G_1, \ldots, G_r$ (that is, $G_i$ is an induced subgraph of $G$ on some $V_i$, and $V = \cup_i V_i$), we have $\ell(G) \leq \sum_{i=1}^{r} \ell(G_i)$.*

In particular, in our case, by combining the above with Proposition 2.1, we have

$$\ell(H) \leq \ell(G) + \ell(\overline{G}) \leq n^{\varepsilon/2} + \lceil \log_2 3n^{\varepsilon/2} \rceil < 3n^{\varepsilon/2}.$$

Finally, label the vertices of $G$ as $\{v_1, \ldots, v_n\}$ and the corresponding vertices of $\overline{G}$ as $\{v'_1, \ldots, v'_n\}$. Following the arguments of the proof of Claim 2.5, it is easy to verify that the set of vertices $\{(v_i, v'_i) : i \in [n]\} \cup \{(v'_i, v_i) : i \in [n]\}$ is an independent set of size $2n$ in $G \times \overline{G} + \overline{G} \times G$, which is an induced subgraph of $H \times H$. Therefore, $c(H) \geq \sqrt{2n}$. ∎

**Remark 3.2:** A standard argument gives a slight improvement in the above lower bound on $c(H)$, to $c(H) \geq 2\sqrt{n}$. See, e.g., [1] (proof of Theorem 2.1) for further details.

## 3.2 Proof of Proposition 1.3

Notice that $\mathrm{minrk}_{p^e}(G) \leq \mathrm{minrk}_{p^d}(G)$ for any prime $p$ and integers $e > d$. Therefore, we can assume without loss of generality that all the $\mathbb{F}_i$-s are fields with pairwise distinct characteristics. Let $G_i$ denote the graph obtained by applying Proposition 2.2 on $\mathbb{K}$ and $\mathbb{F}_i$, so that:

$$\mathrm{minrk}_{\mathbb{K}}(G_i) \leq n^{\varepsilon/2} \text{ and } \mathrm{minrk}_{\mathbb{F}_i}(G_i) \geq n^{1-\varepsilon/2} \text{ ,}$$

and let $G = \sum_{i=1}^{t} G_i$ be the disjoint union of these graphs. Since the adjacency matrix of $G$ is a diagonal block-matrix of the adjacency matrices corresponding to the individual $G_i$-s, we obtain that

$$\mathrm{minrk}_{\mathbb{K}}(G) = \sum_{i=1}^{t} \mathrm{minrk}_{\mathbb{K}}(G_i) \leq tn^{\varepsilon/2} < n^{\varepsilon} \text{ ,}$$

Clearly, for every $i$, $\mathrm{minrk}_{\mathbb{F}_i}(G) \geq \mathrm{minrk}_{\mathbb{F}_i}(G_i)$, completing the proof. ∎

## 4 The problem definition revisited

Call the problem of finding the optimal index code, as defined in Definition 1, **Problem 1**. At first glance, Problem 1 seems to capture only very restricted instances of the source coding problem for ISCOD, and its motivating applications in communication. Namely, the main restrictions are:

*(1) Each receiver requests exactly one data block.*

*(2) Each data block is requested only once.*

*(3) Every data block consists of a single bit.*

In [5], where Definition 1 was stated, it is proved that the source coding problem for ISCOD can be reduced to a similar one which satisfies restriction (1). This is achieved by replacing a user that requests $k > 1$ blocks by $k$ users, all having the same side information, and each requesting a different block. On the other hand, restriction (2) appeared in [5] to simplify the problem and to enable the side-information to be modeled by a directed graph.[3] Restriction (3) is stated assuming a larger block size does not dramatically effect the nature of the problem. In what follows, we aim to reconsider the last two restrictions.

## 4.1 Larger alphabet and multiple rounds

Suppose the data string $x$ is over a possibly larger alphabet, e.g., $\{0,1\}^t$ for some $t \geq 1$:

**Problem 2:** The generalization of Problem 1, where each input symbol $x_i \in \{0,1\}^t$ comprises a *block* of $t$ bits. Every user is interested in a single block, and knows a subset of the other blocks.

By considering each of the $t$ bits of the symbol as one independent round of transmission, one can verify that the following formulation is equivalent:

**Problem 2':** The generalization of Problem 1 to $t \geq 1$ rounds over the same side information graph $G$. The sender wishes to transmit $t$ words $x^1, \ldots, x^t \in \{0,1\}^n$, with the same side information setting. Receiver $R_i$ is always interested in the $i$-th bit of the input words, $x_i^1, \ldots, x_i^t$.

The above problem can be reduced to Problem 1 by considering the graph $G[t]$, defined as follows. For some integer $t$, let $G[t]$ denote the *t-blow-up* of $G$ (with independent sets), that is, the graph on the vertex set $V(G) \times [t]$, where $(u,i)$ and $(v,j)$ are adjacent iff $uv \in E(G)$. Indeed, Problem 2 reduces to Problem 1 with the side information graph $G[t]$, by assigning a receiver to each of the data bits. Therefore, this extension is in fact a special case of the original seemingly restricted problem.

Clearly, one may choose to treat each round of transmission independently, at a total cost of $t \cdot \ell(G)$ transmitted bits, thus $\ell(G[t]) \leq t \cdot \ell(G)$. The next remark shows that this bound is sometimes tight:

**Remark 4.1:** If an undirected graph $G$ satisfies $\ell(G) = \alpha(G)$ (this holds, e.g., for all graphs satisfying $\alpha(G) = \chi(\overline{G})$, and namely for perfect graphs), then $\ell(G[t]) = t \cdot \ell(G)$, as

$$t \cdot \ell(G) \geq \ell(G[t]) \geq \alpha(G[t]) = t \cdot \alpha(G) = t \cdot \ell(G) \ .$$

However, as the next remark states, one may indeed save on communication when sending a unified transmission for the entire set of rounds (or block of symbols):

---

[3]It followed the observation that if the same block is requested by several receivers, then most of the communication saving comes from transmitting this block once (*duplicate elimination*).

**Remark 4.2:** In a subsequent work [2], we show that there are graphs for which $\ell(G[t]) < t \cdot \ell(G)$. That is, transmission of $t$ rounds may strictly improve upon the performance of $t$ independent transmissions. This justifies the study of the *index coding rate* defined by

$$\lim_{t \to \infty} \frac{\ell(G[t])}{t}$$

(the limit exists by sub-additivity). This corresponds to the average length of a codeword per round, when the number of rounds tends to infinity.

A natural extension of Problem 2' is the case where the underlying side information graph changes between rounds:

**Problem 3:** The generalization of Problem 1 to $t \geq 1$ rounds: the sender wishes to transmit $t$ words $x^1, \ldots, x^t \in \{0,1\}^n$, with respective side information graphs $G_1, \ldots, G_t$. Receiver $R_i$ is always interested in the $i$-th bit of the input words, $x_i^1, \ldots, x_i^t$.

Even in this more general setting, a reduction to Problem 1 is possible: let $G = G_1 \circ \cdots \circ G_t$ denote the directed graph on the vertex set $V(G) = [n] \times [t]$, where for all $i_1, i_2 \in [n]$ and $k_1, k_2 \in [t]$, $((i_1, k_1), (i_2, k_2))$ is an edge of $G$ iff $(i_1, i_2) \in E(G_{k_2})$. Again, it is straightforward to see that $\ell(G)$ is precisely the solution for Problem 3.

In the general setting of Problem 3, it is even simpler to see that independent transmissions may consume significantly more communication. For instance, consider the following case. We have two receivers, $R_1$ and $R_2$, and two rounds for transmitting the binary words $x = x_1 x_2$ and $y = y_1 y_2$. Suppose that in the first round receiver $R_1$ knows $x_2$ and in the second transmission receiver $R_2$ knows $y_1$. In this case, each round - if transmitted separately - requires 2 bits to be transmitted. Yet, if the server transmits the 3 bits

$$x_1 \oplus y_1 \ , \ x_2 \oplus y_2 \ , \ x_1 \oplus y_2 \ ,$$

then both receivers can reconstruct their missing bits (and moreover, reconstruct all of $x$ and $y$).

This in fact is a special case of the following construction. We define a pair of graphs $G_1, G_2$ such that $\ell(G_1) = \ell(G_2) = n$ and yet only $\ell(G_1 \circ G_2) = n + 1$ bits need to transmitted for consecutive transmissions. This is stated in the next claim, where the *transitive tournament graph* on $n$ vertices is isomorphic to the directed graph on the vertex set $[n]$, where $(i, j)$ is an edge iff $i < j$.

**Claim 4.3.** *Let $G_1$ denote the transitive tournament graph on $n$ vertices, and let $G_2$ denote the graph obtained from $G_1$ by reversing all edges. Then $\ell(G_1) + \ell(G_2) = 2n$, and yet $\ell(G_1 \circ G_2) = n + 1$.*

*Proof.* Without loss of generality, assume that $E(G_1) = \{(i, j) : i < j\}$ and $E(G_2) = \{(i, j) : i > j\}$. Since $G_1$ and $G_2$ are both acyclic, the fact that $\ell(G_1) = \ell(G_2) = n$ follows from the lower bound of [4] ($\ell(G)$ is always at least the size a maximum induced acyclic subgraph of $G$).

Recall that by definition, $G_1 \circ G_2$ is the disjoint union of $G_1$ and $G_2$, with the additional edges $\{((i, 1), (j, 2)) : j < i\}$ and $\{((i, 2), (j, 1)) : j > i\}$. Therefore, $G_1 \circ G_2$ has an induced acyclic graph of size $n + 1$: for instance, the set $\{(i, 1) : i \in [n]\} \cup \{(n, 2)\}$ induces such a graph. We deduce that $\ell(G_1 \circ G_2) \geq n + 1$.

To complete the proof of the claim, we give an encoding scheme for $G_1 \circ G_2$ which requires the transmission of $n+1$ bits, hence $\ell(G_1 \circ G_2) \le n+1$. Denote the two words to be transmitted by $x = x_1 \ldots x_n$ and $y = y_1 \ldots y_n$. The coding scheme is linear: by transmitting $x_i \oplus y_i$ for $i \in [n]$ and $\oplus_{i \in [n]} x_i$, it is not difficult to see that each receiver is able to decode its missing bits (in fact, each receiver can reconstruct all the bits of $x$ and $y$). ∎

## 4.2 Shared requests

**Problem 4:** The generalization of Problem 1 to $m \ge n$ receivers, each interested in a single bit (i.e., we allow several users to ask for the same bit).

In this case, the one-to-one correspondence between message bits and receivers no longer holds, thus the directed side information graph seems unsuitable. However, it is still possible to obtain bounds on the optimal linear and non-linear codes using slightly different models.

Let $\mathcal{P}_4$ denote an instance of Problem 4, and let $\ell(\mathcal{P}_4)$ denote the length of an optimal index code in this setting. It is convenient to model the side-information of $\mathcal{P}_2$ using a binary $m \times n$ matrix, where the $ij$ entry is 1 iff the $i$-th user knows the $j$-th bit (if $m = n$, this matrix is the adjacency matrix of the side information graph). With this in mind, we extend the notion of representing the side-information graph as follows: an $m \times n$ matrix $B$ *represents* $\mathcal{P}_4$ over $\mathbb{F}$ iff for all $i$ and $j$:

- If the $i$-th receiver is interested in the bit $x_j$, then $B_{ij} \ne 0$.

- If the $i$-th receiver is neither interested in nor knows the bit $x_j$, then $B_{ij} = 0$.

Notice that in the special case $m = n$, the above definition coincides with the usual definition of representing the side-information graph. Let $\text{minrk}_{\mathbb{F}}(\mathcal{P}_4)$ denote the minimum rank of a matrix $B$ that represents $\mathcal{P}_4$ over $\mathbb{F}$. It is straightforward to verify that results analogous to Proposition 2.1 and Remark 2.3 hold for this extended notion of matrix representation:

**Theorem 4.4.** *Let $\mathcal{P}_4$ denote an instance of Problem 4. Then the length of an optimal linear code is $\text{minrk}_2(\mathcal{P}_4)$, and the upper bounds of Theorem 2.1 on arbitrary index codes hold for $\mathcal{P}_4$ as well.*

Next, given $\mathcal{P}_4$, define the following two directed $m$-vertex graphs $G_{\text{ind}}$ and $G_{\text{cl}}$. Both vertex sets correspond to the $m$ users, where each set of users interested in the same bit forms an independent set in $G_{\text{ind}}$ and a clique in $G_{\text{cl}}$. In the remaining cases, in both graphs $(v_i, v_j)$ is an edge iff the $i$-th user knows the bit in which the $j$-th user is interested (for $m = n$, both graphs are equal to the usual side-information graph defined in Definition 2). The following simple claim provides additional bounds on $\ell(\mathcal{P}_4)$; we omit the details of its proof.

**Claim 4.5.** *If $\mathcal{P}_4$ denotes an instance of Problem 4, and $G_{\text{ind}}$ and $G_{\text{cl}}$ are defined as above, then:*

1. *$\ell(G_{\text{cl}}) \le \ell(\mathcal{P}_4)$, and in addition, $\text{minrk}_{\mathbb{F}}(G_{\text{cl}}) \le \text{minrk}_{\mathbb{F}}(\mathcal{P}_4)$ for all $\mathbb{F}$.*

2. *$\ell(\mathcal{P}_4) \le \ell(G_{\text{ind}})$, and in addition, $\text{minrk}_{\mathbb{F}}(\mathcal{P}_4) \le \text{minrk}_{\mathbb{F}}(G_{\text{ind}})$ for all $\mathbb{F}$.*

# 5    Concluding remarks and open problems

- In this paper we have introduced constructions of graphs for which linear index coding is suboptimal (Theorem 1.1), thus disproving the main conjecture of [4]. It is in fact shown that any linear index code for these $n$-vertex graphs requires a word length of $n^{1-o(1)}$ bits (barely improving the naïve protocol which requires $n$ bits), yet a given index code for these graphs utilizes words which are only $n^{o(1)}$ bits long.

- The graphs constructed extend Alon's variant of the Ramsey construction given by Frankl and Wilson. For these graphs, linear index codes over higher-order fields outperform the linear codes over $GF(2)$. Furthermore, a variant of this construction (given in Theorem 1.2) shows that there are graphs where linear codes over *any* field are suboptimal.

- The main question for further work is trying to obtain tight bounds on $\ell(G)$ for a general graph $G$. In addition, it would be interesting to determine the expected value of $\ell(G)$ for the random graph $G \sim \mathcal{G}(n, \frac{1}{2})$.

- In Theorem 1.1, we have constructed $n$-vertex graphs, where the ratio between the parameters $\mathrm{minrk}_2(G)$ and $\ell(G)$ was $n/\exp(O(\sqrt{\log n \log \log n}))$. It can be interesting to obtain an even larger gap between these two parameters, and namely, to show $n$-vertex graphs $G$ where $\mathrm{minrk}_2(G)/\ell(G) = \widetilde{\Theta}(n)$. This may require either a different approach to the problem, or significantly improving the given Ramsey constructions.

- In addition, we showed that more general scenarios of index coding, as presented in [5], can be reduced to the main problem, which recently attracted attention. In this context, we have demonstrated that one may save on communication when transmitting $t$ binary words at once, rather than transmitting these words independently. We have shown this for the case where the underlying side-information graph is allowed to change dynamically.

- The most interesting scenario is that of large data blocks over a fixed side-information graph. As in [4], we have confined ourselves in this paper mainly to the case in which each of the data blocks consists of a single bit. However, this analysis of index coding is relevant to the motivating application only if the communication which is required to coordinate the side information graph is negligible with respect to the size of the data blocks themselves. Therefore, we should, in fact, consider a scenario in which an $n$-word of $b$-bits blocks is transmitted, where $b \gg n$. In this case, it is clearly possible to use an optimal index code for each bit in the block independently, transmitting $b \cdot \ell(G)$ bits altogether. Nevertheless, this protocol is not guaranteed to be optimal, which yields the following natural question:

  Is there a side information graph $G$ on $n$ vertices and integer $b$, for which transmitting an $n$-word which consists of $b$-bits blocks requires less than $b \cdot \ell(G)$ bits?

**Remark 5.1:**  After the completion of this work, with Noga Alon, we were able to answer the last question in the affirmative. This will appear in a subsequent work [2].

**Acknowledgement:** We are grateful to Noga Alon and Oded Regev for helpful discussions. We would also like to thank the FOCS 2007 program committee for helpful suggestions.